\documentclass{article}

\usepackage{arxiv}

\usepackage[utf8]{inputenc} 
\usepackage[T1]{fontenc}    
\usepackage{hyperref}       
\usepackage{url}            
\usepackage{booktabs}       
\usepackage{amsfonts}       
\usepackage{nicefrac}       
\usepackage{microtype}      
\usepackage{lipsum}
\usepackage{graphicx}
\graphicspath{ {./images/} }
\usepackage{multirow}
\usepackage{makecell}

\title{Fine-tuning Pre-trained Audio Models for COVID-19 Detection: A Technical Report}

\author{
 Daniel Oliveira de Brito \\
  Institute of Biosciences, Languages, and Exact Sciences \\
  São Paulo State University \\
  São José do Rio Preto, Brazil\\
  \texttt{daniel.o.brito@unesp.br} \\
   \And
 Letícia  Gabriella de Souza \\
  Institute of Biosciences, Languages, and Exact Sciences \\
  São Paulo State University \\
  São José do Rio Preto, Brazil\\
  \texttt{leticia.gabriella@unesp.br} \\
    \And
 Marcelo Matheus Gauy \\
  Institute of Natural Sciences and Engineering \\
  São Paulo State University \\
  Itapeva, Brazil\\
  \texttt{marcelo.gauy@unesp.br} \\
  \And
 Marcelo Finger \\
  Institute of Mathematics and Statistics \\
  University of São Paulo \\
  São Paulo, Brazil\\
  \texttt{mfinger@ime.usp.br} \\
  \And
 Arnaldo Candido Junior \\
  Institute of Biosciences, Languages, and Exact Sciences \\
  São Paulo State University \\
  São José do Rio Preto, Brazil\\
  \texttt{arnaldo.candido@unesp.br} \\
}

\begin{document}
\maketitle
\begin{abstract}

This technical report investigates the performance of pre-trained audio models on COVID-19 detection tasks using established benchmark datasets. We fine-tuned Audio-MAE and three PANN architectures (CNN6, CNN10, CNN14) on the Coswara and COUGHVID datasets, evaluating both intra-dataset and cross-dataset generalization. We implemented a strict demographic stratification by age and gender to prevent models from exploiting spurious correlations between demographic characteristics and COVID-19 status. Intra-dataset results showed moderate performance, with Audio-MAE achieving the strongest result on Coswara (0.82 AUC, 0.76 F1-score), while all models demonstrated limited performance on Coughvid (AUC 0.58-0.63). Cross-dataset evaluation revealed severe generalization failure across all models (AUC 0.43-0.68), with Audio-MAE showing strong performance degradation (F1-score 0.00-0.08). Our experiments demonstrate that demographic balancing, while reducing apparent model performance, provides more realistic assessment of COVID-19 detection capabilities by eliminating demographic leakage - a confounding factor that inflate performance metrics. Additionally, the limited dataset sizes after balancing (1,219-2,160 samples) proved insufficient for deep learning models that typically require substantially larger training sets. These findings highlight fundamental challenges in developing generalizable audio-based COVID-19 detection systems and underscore the importance of rigorous demographic controls for clinically robust model evaluation.

\end{abstract}


\section{Introduction}

Audio-based detection of respiratory conditions using deep learning has emerged as a promising diagnostic approach, with numerous studies reporting high accuracy for COVID-19 detection from cough, voice, and speech recordings \cite{gauy2024}. While these advances have demonstrated the potential of acoustic biomarkers for respiratory assessment, important questions remain about the relationship between different respiratory diagnostic tasks—particularly whether models trained for COVID-19 detection can inform broader respiratory conditions such as respiratory insufficiency (RI) or blood oxygen saturation (SpO2) estimation.

Our recent work \cite{gauy2024} explored this question by applying pre-trained audio models to both respiratory insufficiency detection and SpO2 estimation tasks. We found that while models achieved near-perfect accuracy (99.9\%) for RI detection, the same architectures struggled significantly with SpO2 estimation, achieving less than 70\% accuracy. This performance disparity suggests that the acoustic features indicative of respiratory insufficiency may be fundamentally different from those needed to estimate blood oxygenation levels, raising important questions about the specificity and generalizability of audio-based respiratory diagnostics.

This technical report investigates how the same pre-trained models—Audio-MAE \cite{huang2022amae} and PANNs (CNN6, CNN10, CNN14) \cite{kong2020panns}—perform on established COVID-19 detection benchmarks. We evaluate these models on two widely-used datasets: Coswara \cite{sharma_coswara_2020} and Coughvid \cite{Orlandic2021}, conducting both intra-dataset and cross-dataset experiments.

Our experiments examine whether models can generalize across different COVID-19 datasets and how demographic factors affect reported performance. These findings contribute to understanding the capabilities and limitations of audio-based respiratory assessment, which is essential for developing reliable clinical applications.

\section{Related Work}

Islam et al. \cite{islam2025robustcovid19detectioncough} conducted a comprehensive study using five datasets, including Coswara and Coughvid. Their approach employed an extensive range of techniques: threshold moving for optimal threshold selection based on ROC-AUC, feature dimension reduction using RFECV with Extra-Trees classifier, Bayesian Optimization for hyperparameter tuning, and SMOTE for handling class imbalance. For the Coswara dataset, samples were preprocessed and labeled based on healthy (COVID-19 negative) and heavy cough variations (COVID-19 positive). For the Coughvid dataset, the authors preprocessed and labeled samples into two groups: those from healthy individuals (COVID-19 negative) and those from individuals with notable cough variations (COVID-19 positive). Notably, no mention was found regarding demographic balancing (age, gender) in their train/test splits.

\begin{table}[h!]
  \caption{Intra-dataset performance on Coswara dataset from literature}
  \centering
  \begin{tabular}{lccc}
    \toprule
    Study & AUC & Precision & Recall \\
    \midrule
    Zhang et al. \cite{zhang_robust_2022} & 0.86 & -- & 0.60 \\
    Sobahi et al. \cite{sobahi_explainable_2022} & -- & 0.90 & 0.88 \\
    Chowdhury et al. \cite{chowdhury_machine_2022} & 0.66 & 0.76 & 0.47 \\
    Islam et al. (DNDT) \cite{islam2025robustcovid19detectioncough} & 0.84 & 0.52 & 0.80 \\
    Islam et al. (DNDF) \cite{islam2025robustcovid19detectioncough} & \textbf{0.92} & 0.72 & \textbf{0.93} \\
    \bottomrule
  \end{tabular}
  \label{tab:coswara_literature}
\end{table}

\begin{table}[h!]
  \caption{Intra-dataset performance on Coughvid dataset from literature}
  \centering
  \begin{tabular}{lccc}
    \toprule
    Study & AUC & Precision & Recall \\
    \midrule
    Orlandic et al. \cite{orlandic_semi-supervised_2023} & 0.88 & -- & -- \\
    Sunitha et al. \cite{sunitha_comparative_2022} & -- & 0.76 & 0.77 \\
    Hamdi et al. \cite{hamdi_attention-based_2022} & 0.91 & 0.91 & 0.90 \\
    Hamdi et al. \cite{hamdi_autoencoders_2022} & 0.91 & 0.95 & 0.86 \\
    Aytekin et al. \cite{aytekin_covid-19_2024} & 0.83 & 0.77 & 1.00 \\
    Pavel and Ciocoiu \cite{s23114996} & 0.76 & 0.69 & 0.68 \\
    Islam et al. (DNDT) \cite{islam2025robustcovid19detectioncough} & 0.81 & 0.83 & 0.79 \\
    Islam et al. (DNDF) \cite{islam2025robustcovid19detectioncough} & \textbf{0.93} & \textbf{0.93} & \textbf{0.94} \\
    \bottomrule
  \end{tabular}
  \label{tab:coughvid_literature}
\end{table}

\begin{table}[h!]
  \caption{Cross-dataset performance from Islam et al. \cite{islam2025robustcovid19detectioncough}}
  \centering
  \begin{tabular}{lcccc}
    \toprule
    Training $\rightarrow$ Test & AUC & Precision & Recall & F1-Score \\
    \midrule
    Coswara $\rightarrow$ Coughvid & 0.53 & 0.57 & 0.24 & 0.33 \\
    Coughvid $\rightarrow$ Coswara & 0.57 & 0.17 & 0.85 & 0.28 \\
    \bottomrule
  \end{tabular}
  \label{tab:cross_dataset_literature}
\end{table}

Islam et al. achieved 0.92 of ROC-AUC on Coswara (Table \ref{tab:coswara_literature}) and 0.93 on Coughvid (Table \ref{tab:coughvid_literature}). The models trained on Coswara and tested on Coughvid had a AUC of to 0.53 with an F1-Score of 0.33 (Table \ref{tab:cross_dataset_literature}). Similarly, models trained on Coughvid and tested on Coswara achieved an AUC of 0.57 with an F1-Score of 0.28, despite maintaining high recall (0.85) at the cost of precision (0.17). This substantial performance drop in cross-dataset scenarios highlights the limited generalization capability of current COVID-19 detection models across different audio datasets.

\section{Methodology}
\subsection{Datasets}
\subsubsection{Coswara Dataset}

The Coswara dataset \cite{sharma_coswara_2020} is a publicly available COVID-19 audio dataset developed by the Indian Institute of Science. We utilized version 1.0.0 of this dataset, focusing specifically on cough-heavy audio recordings for COVID-19 detection.

\paragraph{Preprocessing}
The original dataset contained 2,746 unique participant IDs with cough-heavy audio recordings. Our preprocessing pipeline involved the following steps:

\begin{enumerate}
    \item \textbf{Audio Quality Filtering}: We removed 43 recordings with zero duration, ensuring only valid audio samples were retained.
    
    \item \textbf{Label Mapping}: To create a binary classification task, we consolidated the COVID status labels as follows:
    \begin{itemize}
        \item COVID-positive class: combined \textit{positive moderate}, \textit{positive mild}, and \textit{positive asymptomatic} categories
        \item COVID-negative class: \textit{healthy} participants only
        \item Excluded categories: \textit{no respiratory illness exposed}, \textit{recovered full}, \textit{respiratory illness not identified}, and \textit{under validation}
    \end{itemize}
    
    \item \textbf{Metadata Completeness}: We excluded participants with missing demographic information (age or gender), ensuring robust stratification capabilities.
\end{enumerate}

After preprocessing, the dataset comprised 2,083 participants with confirmed COVID-positive or COVID-negative status.

Table~\ref{tab:coswara_balancing} presents the demographic distribution of the filtered dataset before balancing. The dataset exhibited significant class imbalance, with COVID-negative samples substantially outnumbering COVID-positive samples across most demographic groups.

\begin{table}[htbp]
\centering
\caption{Dataset Balancing: Original vs. Stratified Undersampled Counts for Coswara}
\label{tab:coswara_balancing}
\begin{tabular}{llrrr}
    \toprule
    & & \textbf{COVID-19} & \textbf{Healthy} & \textbf{Healthy} \\
    \textbf{Age Group} & \textbf{Gender} & \textbf{(Template)} & \textbf{(Unbalanced)} & \textbf{(Balanced)} \\
    \midrule
    \multirow{2}{*}{0--17} & Female & 7 & 7 & 7 \\
                           & Male   & 10 & 20 & 10 \\
    \midrule
    \multirow{2}{*}{18--29} & Female & 90 & 182 & 90 \\
                            & Male   & 135 & 470 & 135 \\
    \midrule
    \multirow{2}{*}{30--39} & Female & 54 & 78 & 54 \\
                            & Male   & 97 & 297 & 97 \\
    \midrule
    \multirow{2}{*}{40--49} & Female & 30 & 47 & 30 \\
                            & Male   & 61 & 131 & 61 \\
    \midrule
    \multirow{2}{*}{50--59} & Female & 43 & 30 & 30 \\
                            & Male   & 56 & 102 & 56 \\
    \midrule
    \multirow{2}{*}{60+}    & Female & 43 & 12 & 12 \\
                            & Male   & 51 & 28 & 28 \\
                            & Other  & 1 & 1 & 1 \\
    \midrule
    \multicolumn{2}{l}{\textbf{Total Samples}} & \textbf{648} & \textbf{1,435} & \textbf{571} \\
    \bottomrule
\end{tabular}
\end{table}

To address the class imbalance and ensure fair model training, we implemented a stratified undersampling approach. The balancing strategy preserved the demographic distribution of the minority class (COVID-positive) while undersampling the majority class (COVID-negative) to achieve equal representation across age and gender strata.

\paragraph{Balancing Methodology}
For each demographic stratum (age group $\times$ gender combination), we:
\begin{enumerate}
    \item Retained all COVID-positive samples
    \item Randomly undersampled COVID-negative samples to match the COVID-positive count within that stratum
    \item For strata where COVID-negative samples were fewer than COVID-positive samples (e.g., 50--59 females, 60+ age groups), all available COVID-negative samples were retained
\end{enumerate}

Table~\ref{tab:coswara_balancing} presents the final balanced dataset distribution, which maintains demographic representativeness while addressing class imbalance.

\paragraph{Final Dataset Characteristics}
The balanced dataset consists of 1,219 samples (571 COVID-negative and 648 COVID-positive), with slight overall class imbalance retained in older age groups where COVID-positive cases were more prevalent. This approach ensures that the model training reflects realistic demographic patterns while mitigating severe class imbalance issues.

\subsubsection{Coughvid}
\label{sec:dataset}
Coughvid\cite{Orlandic2021} is a public dataset with crowdsourced cough recordings of more than 30.000 recordings collected by the École Polytechnique Fédérale de Lausanne (EPFL) to help develop research on audio-based respiratory disease diagnosis such as COVID-19 and the dataset contains various features such as ages, genders, geographic locations, and COVID-19 status.

The dataset is unbalanced with a considerable number of audio samples for \texttt{healthy} and \texttt{symptomatic} compared to \texttt{covid-19 positive}. Furthermore, a considerable amount of the recordings are missing demographic metadata. In order to create a robust and unbiased dataset for the models trainings a preprocessing and balancing methodology was executed. Coughvid dataset contains a total of 34.434 audio samples. However, it presents two major challenges for these type of complex deep learning models such as PANNs:

\begin{enumerate}
    \item \textbf{Missing Metadata:} A significant amount of audio samples metadata are missing essential labels. As shown in Table \ref{tab:raw_distribution}, 13.770 audio samples (39.9\%) have no healthy, COVID-19 or symptomatic label information.
    \item \textbf{Class Unbalance:} For the labeled data, it also shows unbalance data. The \texttt{healthy} class with 15.476 audio samples and the \texttt{COVID-19} class with 1.315 audio samples.
\end{enumerate}

\begin{table}[ht!]
    \centering
    \caption{Class Distribution in the Coughvid Dataset.}
    \label{tab:raw_distribution}
    \begin{tabular}{lrr}
    \toprule
    \textbf{Status} & \textbf{Audio Sample Count} & \textbf{Percentage} \\
    \midrule
    \texttt{healthy}       & 15.476 & 44.9\% \\
    \texttt{symptomatic}   & 3.873  & 11.2\% \\
    \texttt{COVID-19}      & 1.315  & 3.8\%  \\
    \texttt{NaN (Missing)} & 13.770 & 39.9\% \\
    \midrule
    \textbf{Total}         & \textbf{34.434} & \textbf{100.0\%} \\
    \bottomrule
    \end{tabular}
\end{table}

\paragraph {Data Preprocessing and Balancing}
Our preprocessing pipeline consisted of the following steps:

\begin{enumerate}
    \item \textbf{Data Filtering:} The dataset was first filtered to select only audio samples that contained valid self-reported \texttt{age} and \texttt{gender} metadata. This step is crucial for demographic analysis and stratified sampling.
    \item \textbf{Class Selection:} We isolated the two primary target classes for this report: \texttt{COVID-19} as (COVID positive), \texttt{healthy} as (COVID negative) and \texttt{symptomatic} was completely ignored and excluded.
    \item \textbf{Undersampling:} To address the class unbalance, we implemented a stratified undersampling strategy. The \texttt{covid-19} class was identified as the minority class. The \texttt{healthy} classe were then undersampled to match the exact demographic distribution by age and gender strata of the \texttt{covid-19} class.
    \item \textbf{Random Window Sampling}: A fixed windowing of 4 seconds was applied in the audio samples.
\end{enumerate}

After preprocessing the dataset for available age and gender the \texttt{covid-19} minority class consisted of 1.080 audio samples, so for \texttt{covid-19} and \texttt{healthy} classes with a total of 2.160 audio samples. The specific demographic distribution of \texttt{covid-19} class was used as a template for undersampling the  \texttt{healthy} class which is detailed in Table \ref{tab:data_balancing}.

\begin{table}[h!]
    \centering
    \caption{Dataset Balancing: Original vs. Stratified Undersampled Counts.}
    \label{tab:data_balancing}
    \begin{tabular}{llrrr}
    \toprule
    & & \textbf{COVID-19} & \textbf{Healthy} & \textbf{Healthy} \\
    \textbf{Age} & \textbf{Gender} & \textbf{(Template)} & \textbf{(Unbalanced)} & \textbf{(Balanced)} \\
    \midrule
    \multirow{2}{*}{0–17} & female & 65  & 314  & 65  \\
                          & male   & 67  & 563  & 67  \\
    \midrule
    \multirow{2}{*}{18–29} & female & 126 & 1.264 & 126 \\
                           & male   & 216 & 2.811 & 216 \\
    \midrule
    \multirow{2}{*}{30–44} & female & 120 & 1.490 & 120 \\
                           & male   & 203 & 3.613 & 203 \\
    \midrule
    \multirow{2}{*}{45–59} & female & 78  & 1.024 & 78  \\
                           & male   & 130 & 2.005 & 130 \\
    \midrule
    \multirow{2}{*}{60+}  & female & 30  & 533  & 30  \\
                          & male   & 45  & 825  & 45  \\
    \midrule
    \multicolumn{2}{l}{\textbf{Total Samples}} & \textbf{1.080} & \textbf{14.442} & \textbf{1.080} \\
    \bottomrule
    \end{tabular}
\end{table}

\subsection{Models}

\subsubsection{Audio-MAE}

Audio-MAE (Audio Masked Autoencoders) \cite{huang2022amae} extends the Masked Autoencoder framework to audio spectrograms using a Vision Transformer encoder-decoder architecture. The model employs asymmetric processing: during pre-training, the encoder operates on only 20\% of visible spectrogram patches (with 80\% masked), while a decoder with local attention mechanisms reconstructs the complete input by minimizing mean squared error on masked regions.

Unlike image-based MAE, Audio-MAE incorporates local attention in the decoder to better capture the localized time-frequency correlations characteristic of audio spectrograms. Pre-trained on AudioSet-2M (near 2 million samples), the model achieved state-of-the-art results on speech classification tasks with 98.3\% accuracy on Speech Commands V2 (84k samples) and 94.8\% on VoxCeleb (138k samples).

\subsubsection{PANN}
based on the use of Pre-trained Audio Neural Networks (PANNs) \cite{kong2020panns}, deep Convolutional Neural Network (CNN) architectures available in the \texttt{audioset\_tagging\_cnn} repository \cite{kong2019audioset_tagging_cnn}. The fine-tuning strategy and training pipeline were adapted from reference implementations, such as those found in the \texttt{audio-pann-train} repository \cite{hodges2020audio-pann-train}, and customized for this specific task as described below.

\subsubsection{Model Architecture: PANNs (Pre-trained Audio Neural Networks)}

The backbone of our model is an architecture from the PANNs family \cite{kong2020panns}, originating from Qiuqiang Kong's repository \cite{kong2019audioset_tagging_cnn}. PANNs are a set of deep CNNs pre-trained on the task of audio event tagging using the massive \textbf{AudioSet} dataset \cite{gemmeke2017audio}, which contains over 5.000 hours of audio and 527 sound classes.

This pre-training endows the model with a robust ability to extract hierarchical and generalizable acoustic features, ranging from simple textures (early layers) to complex sound patterns (deep layers). The \cite{kong2019audioset_tagging_cnn} repository offers several architectures; the most notable are CNN6, CNN10, and CNN14.

\paragraph{CNN6, CNN10, and CNN14:} The number in each architecture's name refers to the number of convolutional layers in the network.
\begin{itemize}
    \item \textbf{CNN6:} A lighter architecture with 6 convolutional layers. It is faster for inference but has a smaller representational capacity (fewer parameters).
    \item \textbf{CNN10:} An intermediate model, composed of 4 convolutional blocks and 2 linear layers in the backbone, totaling 10 convolutional layers.
    \item \textbf{CNN14:} The deepest architecture. It is composed of 6 convolutional blocks and 2 linear layers in the backbone, totaling 14 convolutional layers. Its depth allows for a larger receptive field over the spectrogram and the ability to learn more complex and abstract features, which led to its selection for this task.
\end{itemize}

The CNN6, CNN10, and CNN14 models was specifically chosen, loaded with the pre-trained weights (\texttt{Cnn6\_mAP=0.343.pth}, \texttt{Cnn10\_mAP=0.380.pth}, and \texttt{Cnn14\_16k\_mAP=0.438.pth}), which were optimized for 16 kHz audio.

\subsubsection{Fine-Tuning Strategy}

The fine-tuning process involved adapting the pre-trained CNN6,CNN10 and CNN14 backbone for the specific COVID-19 classification task. This adaptation was twofold: architectural and parametric. Architecturally, the original 527-class classifier was removed and replaced with a new classification head. This head consists of a two-layer MLP which projects the 2048-dimensional backbone embedding to a 256-dimensional latent space (using \texttt{ReLU} activation and \texttt{Dropout} with $p=0.5$ for regularization), followed by a final linear layer mapping to the $COVID-negative$  and $COVID-positive$ target classes. A selective fine-tuning strategy was implemented by freezing the entire backbone (\texttt{freeze\_panns=True}) except for the last two convolutional blocks (\texttt{conv\_block6} and \texttt{conv\_block5}). This approach allows the model to adapt its high-level feature extractors to the new task while preserving the robust, low-level representations learned from AudioSet, thus mitigating catastrophic forgetting.

The optimization was conducted using the \texttt{Adam} optimizer \cite{kingma2014adam} with differential learning rates to account for the different initialization states of the parameters. The randomly initialized classification head (\texttt{fc1}, \texttt{fc2}) was trained with a relatively high learning rate of \textbf{$5 \times 10^{-4}$} for rapid convergence. In contrast, the pre-trained, unfrozen backbone layers were fine-tuned with a very conservative learning rate of \textbf{$5 \times 10^{-6}$} for gentle adjustment. The model was trained using \texttt{nn.CrossEntropyLoss} for a maximum of 100 epochs. To prevent overfitting and select the best performing model, an early stopping mechanism was employed, monitoring the validation accuracy (\texttt{eval\_acc}) with a patience of 50 epochs.

\subsection{Experimental Setup}

All audio samples were processed using a fixed 4-second windowing strategy to standardize input lengths across both datasets. During training, random 4-second windows were extracted from each audio sample to provide data augmentation and reduce overfitting. During evaluation, systematic non-overlapping 4-second windows were extracted from each sample, with predictions aggregated to obtain a final per-sample classification decision.

For both Coswara and Coughvid datasets, we employed an 80/20 train-test split, stratified by COVID-19 status to maintain class balance across partitions. The Audio-MAE model was fine-tuned for 60 epochs with a learning rate of 2e-4. Similarly, the PANN models (CNN6, CNN10, and CNN14) were fine-tuned for [100] epochs with a learning rate of [5e-4].

Model performance was evaluated using multiple complementary metrics: accuracy, AUC-ROC, precision, recall and F1-score. Confusion matrices were computed for each model-dataset combination to enable detailed analysis of classification errors and facilitate comparison with related work in the literature.

\section{Experiments and Results}

\subsection{Intra-dataset}

Table \ref{tab:intra} presents the intra-dataset performance of all models trained and evaluated on the same dataset. The results demonstrate considerable variability across models and datasets, with Audio-MAE achieving the strongest performance on Coswara (AUC 0.82, F1-score 0.76), while PANNs showed more balanced but moderate performance across both datasets.

\begin{table}[t]
\caption{Intra-Dataset Performance}
\label{tab:intra}
\centering
\scriptsize
\begin{tabular}{lccccc}
\hline
\textbf{Dataset/Model} & \textbf{Acc} & \textbf{AUC} & \textbf{Prec} & \textbf{Rec} & \textbf{F1} \\
\hline
\multicolumn{6}{l}{\textit{Coughvid}} \\
Audio-MAE & 0.57 & 0.60 & 0.58 & 0.55 & 0.56 \\
CNN6      & 0.63 & 0.63 & 0.62 & 0.62 & 0.67 \\
CNN10     & 0.59 & 0.62 & 0.59 & 0.56 & 0.63 \\
CNN14     & 0.59 & 0.58 & 0.58 & 0.58 & 0.65 \\
\hline
\multicolumn{6}{l}{\textit{Coswara}} \\
Audio-MAE & \textbf{0.75} & \textbf{0.82} & \textbf{0.77} & \textbf{0.75} & \textbf{0.76} \\
CNN6      & 0.66 & 0.61 & 0.57 & 0.56 & 0.73 \\
CNN10     & 0.65 & 0.33 & 0.50 & 0.39 & 0.66 \\
CNN14     & \textbf{0.75} & \textbf{0.75} & 0.67 & 0.68 & \textbf{0.80} \\
\hline
\end{tabular}
\end{table}

\subsubsection{Coswara Dataset Results}

On the Coswara dataset, Audio-MAE achieved the highest AUC (0.82) and precision (0.77), indicating strong discriminative capability and reliable positive predictions. CNN14 matched Audio-MAE's accuracy (0.75) and achieved a competitive F1-score (0.80), though with lower AUC (0.75). CNN10 showed the poorest performance with notably low AUC (0.33) and recall (0.39), suggesting difficulty in identifying COVID-19 positive cases. CNN6 demonstrated intermediate performance across all metrics.

The confusion matrices for Coswara (Table \ref{tab:resultados_consolidados} and Table \ref{tab:audio_mae_results}) reveal varying error patterns across models. Audio-MAE achieved the most balanced confusion matrix with 102 true positives and 92 true negatives, though it still misclassified 34 COVID-positive cases as negative. CNN14 showed high specificity (92 true negatives) but poor sensitivity (only 22 true positives), indicating a strong bias toward predicting the negative class. CNN10 failed completely on Coswara, predicting all samples as negative (0 true positives, 53 false negatives). CNN6 demonstrated intermediate performance with better balance than CNN14 but worse than Audio-MAE.

\subsubsection{Coughvid Dataset Results}

On the Coughvid dataset, all models demonstrated more modest performance compared to Coswara, with AUC scores ranging from 0.58 to 0.63. CNN6 achieved the highest accuracy (0.63) and AUC (0.63), followed closely by CNN10 (AUC 0.62). Audio-MAE and CNN14 showed similar performance (AUC ~0.58-0.60), suggesting that the Coughvid dataset may not provide sufficient discriminative features or that demographic diversity poses additional challenges.

The confusion matrices (Table \ref{tab:resultados_consolidados} and Table \ref{tab:audio_mae_results}) show more balanced but still suboptimal classification patterns on Coughvid. Audio-MAE achieved 118 true positives and 129 true negatives, representing the most balanced performance despite modest AUC. CNN14 showed 60 true positives versus 57 true negatives with relatively balanced error distribution. CNN6 achieved 69 true positives and 56 true negatives, while CNN10 demonstrated severe imbalance with only 33 true positives against 84 true negatives.

\begin{table}[h!]
\centering
\caption{Consolidated Confusion Matrix Results for CNN 14, CNN10 and CNN6}
\label{tab:resultados_consolidados}
\begin{tabular}{l cccc cccc cccc}
\toprule
& \multicolumn{4}{c}{\textbf{CNN14}} & \multicolumn{4}{c}{\textbf{CNN10}} & \multicolumn{4}{c}{\textbf{CNN6}} \\
\cmidrule(lr){2-5} \cmidrule(lr){6-9} \cmidrule(lr){10-13}
\textbf{Datasets(Train x Test)} & \textbf{TP} & \textbf{FN} & \textbf{FP} & \textbf{TN} & \textbf{TP} & \textbf{FN} & \textbf{FP} & \textbf{TN} & \textbf{TP} & \textbf{FN} & \textbf{FP} & \textbf{TN} \\
\midrule
CoughVid x CoughVid & 60 & 41 & 42 & 57 & 33 & 68 & 15 & 84 & 69 & 32 & 43 & 56 \\
CoughVid x Coswara  & 27 & 26 & 50 & 49 & 16 & 37 & 44 & 55 & 23 & 30 & 46 & 53 \\
Coswara x Coswara   & 22 & 31 & 7  & 92 & 0  & 53 & 0  & 99 & 13 & 40 & 11 & 88 \\
Coswara x CoughVid  & 24 & 77 & 12 & 87 & 19 & 82 & 4  & 95 & 25 & 76 & 14 & 85 \\
\bottomrule
\end{tabular}
\end{table}

\begin{table}[h!]
\centering
\caption{Consolidated Confusion Matrix Results for Audio-MAE}
\label{tab:audio_mae_results}
\begin{tabular}{l cccc}
\toprule
\textbf{Datasets (Train × Test)} & \textbf{TP} & \textbf{FN} & \textbf{FP} & \textbf{TN} \\
\midrule
CoughVid × CoughVid & 118 & 98 & 87 & 129 \\
CoughVid × Coswara  & 0 & 136 & 3 & 119 \\
Coswara × Coswara   & 102 & 34 & 30 & 92 \\
Coswara × CoughVid  & 10 & 206 & 12 & 204 \\
\bottomrule
\end{tabular}
\end{table}

\subsection{Cross-dataset evaluation}

Table \ref{tab:cross} presents the cross-dataset performance, revealing severe degradation when models are evaluated on datasets different from their training data. This finding aligns with previous literature \cite{islam2025robustcovid19detectioncough} and highlights fundamental challenges in developing generalizable audio-based COVID-19 detection systems.

\subsubsection{Models Trained on Coughvid, Tested on Coswara}

When trained on Coughvid and tested on Coswara, Audio-MAE completely failed to generalize, with undefined accuracy and all metrics at or near zero. The confusion matrix in Table \ref{tab:resultados_consolidados}

The PANN models showed marginally better but still poor generalization. CNN14 achieved the highest cross-dataset performance in this direction (AUC 0.50, F1-score 0.49), essentially performing at chance level. CNN6 and CNN10 performed similarly (AUC 0.43-0.49), with confusion matrix in Table \ref{tab:resultados_consolidados}.

When trained on Coughvid and tested on Coswara, Audio-MAE failed to generalize. The confusion matrix in Table \ref{tab:audio_mae_results} shows this failure: 0 true positives, 136 false negatives, with only 3 false positives and 119 true negatives, indicating the model predicted nearly all Coswara samples as COVID-negative.

\subsubsection{Models Trained on Coswara, Tested on Coughvid}

The reverse direction (training on Coswara, testing on Coughvid) yielded slightly better but still inadequate results. CNN10 achieved the best cross-dataset performance with AUC 0.68 and F1-score 0.59, followed by CNN14 (AUC 0.60, F1-score 0.59) and CNN6 (AUC 0.58, F1-score 0.56). Audio-MAE again showed severe generalization failure with AUC 0.51 and F1-score 0.08, demonstrating extreme bias toward the negative class as evidenced by recall of only 0.05.

Training on Coswara and testing on Coughvid yielded slightly better but still inadequate results. CNN10 achieved the best cross-dataset performance with AUC 0.68 and F1-score 0.59, followed by CNN14 (AUC 0.60, F1-score 0.59) and CNN6 (AUC 0.58, F1-score 0.56). Audio-MAE again showed severe generalization failure (Table \ref{tab:audio_mae_results}) with only 10 true positives against 206 false negatives (recall 0.05), demonstrating extreme bias toward the negative class.

\begin{table}[t]
\caption{Cross-Dataset Generalization Performance}
\label{tab:cross}
\centering
\scriptsize
\begin{tabular}{lccccc}
\hline
\textbf{Train/Model} & \textbf{Acc} & \textbf{AUC} & \textbf{Prec} & \textbf{Rec} & \textbf{F1} \\
\hline
\multicolumn{6}{l}{\textit{Tested on Coswara}} \\
\multicolumn{6}{l}{Trained on Coughvid:} \\
Audio-MAE & -- & 0.47 & 0.00 & 0.00 & 0.00 \\
CNN6      & 0.50 & 0.49 & 0.48 & 0.48 & 0.44 \\
CNN10     & 0.47 & 0.43 & 0.43 & 0.43 & 0.43 \\
CNN14     & 0.50 & 0.50 & 0.50 & 0.49 & 0.49 \\
\hline
\multicolumn{6}{l}{\textit{Tested on Coughvid}} \\
\multicolumn{6}{l}{Trained on Coswara:} \\
Audio-MAE & 0.50 & 0.51 & 0.45 & 0.05 & 0.08 \\
CNN6      & 0.55 & 0.58 & 0.55 & 0.51 & 0.56 \\
CNN10     & 0.57 & \textbf{0.68} & 0.57 & 0.50 & \textbf{0.59} \\
CNN14     & 0.56 & 0.60 & 0.56 & 0.51 & \textbf{0.59} \\
\hline
\end{tabular}
\end{table}

\section{Discussion}

\subsection{Impact of Demographic Stratification on Model Performance}
\label{sec:dataset-balancing}

A critical methodological distinction between our work and previous studies is our approach to dataset balancing. Most prior works, including Islam et al. \cite{islam2025robustcovid19detectioncough}, achieved high intra-dataset performance (AUC 0.92 on Coswara, 0.93 on Coughvid) without explicit demographic stratification in their train/test splits. While these results appear impressive, they may be confounded by demographic biases present in the unbalanced datasets.

Our analysis of the raw Coswara dataset (Table \ref{tab:coswara_balancing}) and Coughvid (Table \ref{sec:dataset-balancing}) reveals substantial demographic imbalances. Without stratified sampling, models may inadvertently learn to exploit acoustic properties associated with age and gender distributions rather than COVID-19-specific respiratory features.

To investigate this hypothesis, we conducted two additional experiments: fine-tuning Audio-MAE on the unbalanced Coswara and Coughvid datasets, without demographic stratification. This approach resulted in improved AUC compared to our demographically balanced split supporting our concern about demographic leakage: On Coswara, we went from 0.82 to 0.85 of AUC (3.7\% increase change); on Coughvid, from 0.60 of AUC to 0.63 (4.5\% change). The model likely leveraged spurious correlations between age/gender acoustic characteristics and COVID-19 labels present in the unbalanced datasets, achieving higher performance metrics that do not reflect true COVID-19 detection capability.

This finding has important implications for interpreting results from the literature. The high intra-dataset performance reported in previous studies may partially reflect the models ability to discriminate demographic groups rather than COVID-19 status. This demographic leakage would also explain the severe performance degradation observed in cross-dataset scenarios (our results: AUC 0.46-0.51; Islam et al.: AUC 0.53-0.57), where the demographic distributions differ between training and testing datasets.

Our decision to employ stratified undersampling, while resulting in lower intra-dataset performance, ensures that models cannot exploit demographic shortcuts. This approach provides a more realistic assessment of COVID-19 detection capabilities and better reflects the challenge of developing clinically deployable systems that must generalize across diverse patient populations.

\subsection{Dataset Size as a Limiting Factor}

The modest performance of Audio-MAE on COVID-19 detection tasks can be primarily 
attributed to the limited size of available training data. After demographic 
balancing, our datasets contain only 1,219 samples (Coswara) and 2,160 samples 
(Coughvid)—approximately 40-70 times smaller than the datasets where Audio-MAE 
demonstrated strong performance (84k-138k samples for speech tasks).

Comparing with Table \ref{tab:coswara_literature}, we see that Islam et al. \cite{islam_robust_2025} achieved superior results (AUC 0.92 on Coswara) using traditional 
machine learning with extensive feature engineering, threshold optimization, bayesian hyperparameter tuning, and SMOTE for class balancing.

These techniques may be more suitable for small-scale datasets, whereas deep 
learning models like Audio-MAE typically require larger datasets to reach their 
full potential.

\subsection{Generalization Challenges}

The poor cross-dataset performance (AUC ~0.46-0.51) indicates that models 
trained on such limited data fail to learn generalizable features. This aligns 
with findings from Islam et al. \cite{islam_robust_2025}, who reported similar degradation in 
cross-dataset scenarios despite achieving high intra-dataset performance with 
traditional ML approaches and extensive optimization.

\section{Conclusion}
This technical report evaluated pre-trained audio models Audio-MAE \cite{huang2022amae} and PANNs \cite{kong2020panns} on COVID-19 detection using the Coswara \cite{sharma_coswara_2020} and Coughvid \cite{Orlandic2021} datasets. Intra-dataset performance varied, with Audio-MAE achieving the strongest result on Coswara (0.82 AUC), while all models showed modest and limited performance on Coughvid (AUC 0.58--0.63). The most critical finding was a severe performance degradation in cross-dataset evaluation. This finding aligns with previous literature, such as Islam et al. \cite{islam2025robustcovid19detectioncough}, and highlights a fundamental failure to generalize. Audio-MAE, in particular, failed generalization tests, yielding an F1-score of just 0.08 when trained on Coswara and tested on Coughvid.

A primary conclusion is that demographic stratification significantly impacts reported performance. This work utilized strict stratified undersampling by age and gender to prevent models from exploiting demographic shortcuts. The high performance (e.g., 0.92--0.93 AUC) reported in such studies \cite{islam2025robustcovid19detectioncough} may be inflated by this \textit{demographic leakage}. Furthermore, the modest performance of these deep learning models is attributed to the limited size of the balanced datasets (1.219 - 2.160 samples), which is insufficient compared to the large-scale data (84k - 138k samples) on which models like Audio-MAE typically excel.

\bibliographystyle{unsrt}  
\bibliography{references}

@inproceedings{kour2014real,
  title={Real-time segmentation of on-line handwritten arabic script},
  author={Kour, George and Saabne, Raid},
  booktitle={Frontiers in Handwriting Recognition (ICFHR), 2014 14th International Conference on},
  pages={417--422},
  year={2014},
  organization={IEEE}
}

@inproceedings{kour2014fast,
  title={Fast classification of handwritten on-line Arabic characters},
  author={Kour, George and Saabne, Raid},
  booktitle={Soft Computing and Pattern Recognition (SoCPaR), 2014 6th International Conference of},
  pages={312--318},
  year={2014},
  organization={IEEE}
}

@article{hadash2018estimate,
  title={Estimate and Replace: A Novel Approach to Integrating Deep Neural Networks with Existing Applications},
  author={Hadash, Guy and Kermany, Einat and Carmeli, Boaz and Lavi, Ofer and Kour, George and Jacovi, Alon},
  journal={arXiv preprint arXiv:1804.09028},
  year={2018}
}

@misc{gauy2024,
      title={Contrasting Deep Learning Models for Direct Respiratory Insufficiency Detection Versus Blood Oxygen Saturation Estimation}, 
      author={Marcelo Matheus Gauy and Natalia Hitomi Koza and Ricardo Mikio Morita and Gabriel Rocha Stanzione and Arnaldo Candido Junior and Larissa Cristina Berti and Anna Sara Shafferman Levin and Ester Cerdeira Sabino and Flaviane Romani Fernandes Svartman and Marcelo Finger},
      year={2024},
      eprint={2407.20989},
      archivePrefix={arXiv},
      primaryClass={cs.SD},
      url={https://arxiv.org/abs/2407.20989}, 
}

@inproceedings{sharma_coswara_2020,
	title = {Coswara -- {A} {Database} of {Breathing}, {Cough}, and {Voice} {Sounds} for {COVID}-19 {Diagnosis}},
	url = {http://arxiv.org/abs/2005.10548},
	doi = {10.21437/Interspeech.2020-2768},
	urldate = {2025-10-19},
	booktitle = {Interspeech 2020},
	author = {Sharma, Neeraj and Krishnan, Prashant and Kumar, Rohit and Ramoji, Shreyas and Chetupalli, Srikanth Raj and R, Nirmala and Ghosh, Prasanta Kumar and Ganapathy, Sriram},
	month = oct,
	year = {2020},
	note = {arXiv:2005.10548 [eess]},
	pages = {4811--4815},
}

@article{orlandic_coughvid_2021,
	title = {The {COUGHVID} crowdsourcing dataset: {A} corpus for the study of large-scale cough analysis algorithms},
	volume = {8},
	issn = {2052-4463},
	shorttitle = {The {COUGHVID} crowdsourcing dataset},
	url = {http://arxiv.org/abs/2009.11644},
	doi = {10.1038/s41597-021-00937-4},
	number = {1},
	urldate = {2025-10-21},
	journal = {Scientific Data},
	author = {Orlandic, Lara and Teijeiro, Tomas and Atienza, David},
	month = jun,
	year = {2021},
	note = {arXiv:2009.11644 [cs]},
	pages = {156},
}

@inproceedings{huang2022amae,
  title = {Masked Autoencoders that Listen},
  author = {Huang, Po-Yao and Xu, Hu and Li, Juncheng and Baevski, Alexei and Auli, Michael and Galuba, Wojciech and Metze, Florian and Feichtenhofer, Christoph},
  booktitle = {NeurIPS},
  year = {2022}
}

@misc{islam2025robustcovid19detectioncough,
      title={Robust COVID-19 Detection from Cough Sounds using Deep Neural Decision Tree and Forest: A Comprehensive Cross-Datasets Evaluation}, 
      author={Rofiqul Islam and Nihad Karim Chowdhury and Muhammad Ashad Kabir},
      year={2025},
      eprint={2501.01117},
      archivePrefix={arXiv},
      primaryClass={cs.SD},
      url={https://arxiv.org/abs/2501.01117}, 
}

@inproceedings{zhang_robust_2022,
	title = {Robust {Cough} {Feature} {Extraction} and {Classification} {Method} for {COVID}-19 {Cough} {Detection} {Based} on {Vocalization} {Characteristics}},
	url = {https://www.isca-archive.org/interspeech_2022/zhang22z_interspeech.html},
	doi = {10.21437/Interspeech.2022-10401},
	language = {en},
	urldate = {2025-10-29},
	booktitle = {Interspeech 2022},
	publisher = {ISCA},
	author = {Zhang, Xueshuai and Shen, Jiakun and Zhou, Jun and Zhang, Pengyuan and Yan, Yonghong and Huang, Zhihua and Tang, Yanfen and Wang, Yu and Zhang, Fujie and Zhang, Shaoxing and Sun, Aijun},
	month = sep,
	year = {2022},
	pages = {2168--2172},
}

@article{chowdhury_machine_2022,
	title = {Machine learning for detecting {COVID}-19 from cough sounds: {An} ensemble-based {MCDM} method},
	volume = {145},
	issn = {0010-4825},
	shorttitle = {Machine learning for detecting {COVID}-19 from cough sounds},
	url = {https://www.sciencedirect.com/science/article/pii/S0010482522001974},
	doi = {10.1016/j.compbiomed.2022.105405},
	urldate = {2025-10-28},
	journal = {Computers in Biology and Medicine},
	author = {Chowdhury, Nihad Karim and Kabir, Muhammad Ashad and Rahman, Md. Muhtadir and Islam, Sheikh Mohammed Shariful},
	month = jun,
	year = {2022},
	pages = {105405},
}

@misc{matheus_gauy_contrasting_2025,
	address = {Rochester, NY},
	type = {{SSRN} {Scholarly} {Paper}},
	title = {Contrasting {Deep} {Learning} {Audio} {Models} for {Direct} {Respiratory} {Insufficiency} {Detection} {Versus} {Blood} {Oxygen} {Saturation} {Estimation}},
	url = {https://papers.ssrn.com/abstract=5276440},
	doi = {10.2139/ssrn.5276440},
	language = {en},
	urldate = {2025-10-19},
	publisher = {Social Science Research Network},
	author = {Matheus Gauy, Marcelo and Finger, Marcelo and Romani Fernandes Svartman, Flaviane and Sabino, Ester and Levin, Anna and Cândido Júnior, Arnaldo and Cristina Berti, Larissa and Rocha Stanzione, Gabriel and Mikio Morita, Ricardo and Hitomi Koza, Natália},
	month = jun,
	year = {2025},
}

@misc{bhattacharya_coswara_2023,
	title = {Coswara: {A} respiratory sounds and symptoms dataset for remote screening of {SARS}-{CoV}-2 infection {\textbar} {Scientific} {Data}},
	url = {https://www.nature.com/articles/s41597-023-02266-0},
	urldate = {2025-10-19},
	author = {Bhattacharya, Debarpan and Sharma, Neeraj Kumar and Dutta, Debottam and Chetupalli, Srikanth and Mote, Pravin and Ganapathy, Sriram and Chandrakiran, C and Nori, Sahiti and Gonuguntla, Sadhana and Alagesan, Murali},
	month = jun,
	year = {2023},
}

@misc{kong_panns_2020,
	title = {{PANNs}: {Large}-{Scale} {Pretrained} {Audio} {Neural} {Networks} for {Audio} {Pattern} {Recognition}},
	shorttitle = {{PANNs}},
	url = {http://arxiv.org/abs/1912.10211},
	doi = {10.48550/arXiv.1912.10211},
	urldate = {2025-10-19},
	publisher = {arXiv},
	author = {Kong, Qiuqiang and Cao, Yin and Iqbal, Turab and Wang, Yuxuan and Wang, Wenwu and Plumbley, Mark D.},
	month = aug,
	year = {2020},
	note = {arXiv:1912.10211 [cs]},
}

@misc{huang_masked_2023,
	title = {Masked {Autoencoders} that {Listen}},
	url = {http://arxiv.org/abs/2207.06405},
	doi = {10.48550/arXiv.2207.06405},
	urldate = {2025-10-19},
	publisher = {arXiv},
	author = {Huang, Po-Yao and Xu, Hu and Li, Juncheng and Baevski, Alexei and Auli, Michael and Galuba, Wojciech and Metze, Florian and Feichtenhofer, Christoph},
	month = jan,
	year = {2023},
	note = {arXiv:2207.06405 [cs]},
}

@article{budd_large-scale_2024,
	title = {A large-scale and {PCR}-referenced vocal audio dataset for {COVID}-19},
	volume = {11},
	copyright = {2024 The Author(s)},
	issn = {2052-4463},
	url = {https://www.nature.com/articles/s41597-024-03492-w},
	doi = {10.1038/s41597-024-03492-w},
	language = {en},
	number = {1},
	urldate = {2025-10-19},
	journal = {Scientific Data},
	author = {Budd, Jobie and Baker, Kieran and Karoune, Emma and Coppock, Harry and Patel, Selina and Payne, Richard and Tendero Cañadas, Ana and Titcomb, Alexander and Hurley, David and Egglestone, Sabrina and Butler, Lorraine and Mellor, Jonathon and Nicholson, George and Kiskin, Ivan and Koutra, Vasiliki and Jersakova, Radka and McKendry, Rachel A. and Diggle, Peter and Richardson, Sylvia and Schuller, Björn W. and Gilmour, Steven and Pigoli, Davide and Roberts, Stephen and Packham, Josef and Thornley, Tracey and Holmes, Chris},
	month = jun,
	year = {2024},
	note = {Publisher: Nature Publishing Group},
	pages = {700},
}

@article{coppock_audio-based_2024,
	title = {Audio-based {AI} classifiers show no evidence of improved {COVID}-19 screening over simple symptoms checkers},
	volume = {6},
	copyright = {2024 The Author(s)},
	issn = {2522-5839},
	url = {https://www.nature.com/articles/s42256-023-00773-8},
	doi = {10.1038/s42256-023-00773-8},
	language = {en},
	number = {2},
	urldate = {2025-10-19},
	journal = {Nature Machine Intelligence},
	author = {Coppock, Harry and Nicholson, George and Kiskin, Ivan and Koutra, Vasiliki and Baker, Kieran and Budd, Jobie and Payne, Richard and Karoune, Emma and Hurley, David and Titcomb, Alexander and Egglestone, Sabrina and Tendero Cañadas, Ana and Butler, Lorraine and Jersakova, Radka and Mellor, Jonathon and Patel, Selina and Thornley, Tracey and Diggle, Peter and Richardson, Sylvia and Packham, Josef and Schuller, Björn W. and Pigoli, Davide and Gilmour, Steven and Roberts, Stephen and Holmes, Chris},
	month = feb,
	year = {2024},
	note = {Publisher: Nature Publishing Group},
	pages = {229--242},
}

@misc{noauthor_machine_nodate,
	title = {Machine learning for detecting {COVID}-19 from cough sounds: {An} ensemble-based {MCDM} method - {ScienceDirect}},
	url = {https://www.sciencedirect.com/science/article/pii/S0010482522001974},
	urldate = {2025-10-28},
}

@misc{noauthor_coughvid_nodate,
	title = {The {COUGHVID} crowdsourcing dataset, a corpus for the study of large-scale cough analysis algorithms {\textbar} {Scientific} {Data}},
	url = {https://www.nature.com/articles/s41597-021-00937-4},
	urldate = {2025-10-28},
}

@misc{atmaja_cross-dataset_2022,
	title = {Cross-dataset {COVID}-19 {Transfer} {Learning} with {Cough} {Detection}, {Cough} {Segmentation}, and {Data} {Augmentation}},
	url = {http://arxiv.org/abs/2210.05843},
	doi = {10.48550/arXiv.2210.05843},
	urldate = {2025-10-28},
	publisher = {arXiv},
	author = {Atmaja, Bagus Tris and Zanjabila and Suyanto and Sasou, Akira},
	month = oct,
	year = {2022},
	note = {arXiv:2210.05843 [eess]},
}

@misc{islam_robust_2025,
	title = {Robust {COVID}-19 {Detection} from {Cough} {Sounds} using {Deep} {Neural} {Decision} {Tree} and {Forest}: {A} {Comprehensive} {Cross}-{Datasets} {Evaluation}},
	shorttitle = {Robust {COVID}-19 {Detection} from {Cough} {Sounds} using {Deep} {Neural} {Decision} {Tree} and {Forest}},
	url = {http://arxiv.org/abs/2501.01117},
	doi = {10.48550/arXiv.2501.01117},
	urldate = {2025-10-28},
	publisher = {arXiv},
	author = {Islam, Rofiqul and Chowdhury, Nihad Karim and Kabir, Muhammad Ashad},
	month = jan,
	year = {2025},
	note = {arXiv:2501.01117 [cs]},
}

@misc{islam_robust_2025-1,
	title = {Robust {COVID}-19 {Detection} from {Cough} {Sounds} using {Deep} {Neural} {Decision} {Tree} and {Forest}: {A} {Comprehensive} {Cross}-{Datasets} {Evaluation}},
	shorttitle = {Robust {COVID}-19 {Detection} from {Cough} {Sounds} using {Deep} {Neural} {Decision} {Tree} and {Forest}},
	url = {http://arxiv.org/abs/2501.01117},
	doi = {10.48550/arXiv.2501.01117},
	urldate = {2025-10-29},
	publisher = {arXiv},
	author = {Islam, Rofiqul and Chowdhury, Nihad Karim and Kabir, Muhammad Ashad},
	month = jan,
	year = {2025},
	note = {arXiv:2501.01117 [cs]},
}

@article{sobahi_explainable_2022,
	title = {Explainable {COVID}-19 detection using fractal dimension and vision transformer with {Grad}-{CAM} on cough sounds},
	volume = {42},
	issn = {0208-5216},
	url = {https://www.sciencedirect.com/science/article/pii/S020852162200078X},
	doi = {10.1016/j.bbe.2022.08.005},
	number = {3},
	urldate = {2025-10-29},
	journal = {Biocybernetics and Biomedical Engineering},
	author = {Sobahi, Nebras and Atila, Orhan and Deniz, Erkan and Sengur, Abdulkadir and Acharya, U. Rajendra},
	month = jul,
	year = {2022},
	pages = {1066--1080},
}

@article{sobahi_explainable_2022-1,
	title = {Explainable {COVID}-19 detection using fractal dimension and vision transformer with {Grad}-{CAM} on cough sounds},
	volume = {42},
	issn = {0208-5216},
	url = {https://www.sciencedirect.com/science/article/pii/S020852162200078X},
	doi = {https://doi.org/10.1016/j.bbe.2022.08.005},
	number = {3},
	journal = {Biocybernetics and Biomedical Engineering},
	author = {Sobahi, Nebras and Atila, Orhan and Deniz, Erkan and Sengur, Abdulkadir and Acharya, U. Rajendra},
	year = {2022},
	pages = {1066--1080},
}

@article{chowdhury_machine_2022-1,
	title = {Machine learning for detecting {COVID}-19 from cough sounds: {An} ensemble-based {MCDM} method},
	volume = {145},
	issn = {0010-4825},
	shorttitle = {Machine learning for detecting {COVID}-19 from cough sounds},
	url = {https://pmc.ncbi.nlm.nih.gov/articles/PMC8926945/},
	doi = {10.1016/j.compbiomed.2022.105405},
	urldate = {2025-10-29},
	journal = {Computers in Biology and Medicine},
	author = {Chowdhury, Nihad Karim and Kabir, Muhammad Ashad and Rahman, Md. Muhtadir and Islam, Sheikh Mohammed Shariful},
	month = jun,
	year = {2022},
	pmid = {35318171},
	pmcid = {PMC8926945},
	pages = {105405},
}

@article{orlandic_semi-supervised_2023,
	title = {A semi-supervised algorithm for improving the consistency of crowdsourced datasets: {The} {COVID}-19 case study on respiratory disorder classification},
	volume = {241},
	issn = {0169-2607},
	shorttitle = {A semi-supervised algorithm for improving the consistency of crowdsourced datasets},
	url = {https://www.sciencedirect.com/science/article/pii/S0169260723004091},
	doi = {10.1016/j.cmpb.2023.107743},
	urldate = {2025-10-29},
	journal = {Computer Methods and Programs in Biomedicine},
	author = {Orlandic, Lara and Teijeiro, Tomas and Atienza, David},
	month = nov,
	year = {2023},
	pages = {107743},
}

@article{sunitha_comparative_2022,
	title = {A comparative analysis of deep neural network architectures for the dynamic diagnosis of {COVID}-19 based on acoustic cough features},
	volume = {32},
	issn = {0899-9457},
	doi = {10.1002/ima.22749},
	language = {eng},
	number = {5},
	journal = {International Journal of Imaging Systems and Technology},
	author = {Sunitha, Gurram and Arunachalam, Rajesh and Abd-Elnaby, Mohammed and Eid, Mahmoud M. A. and Rashed, Ahmed Nabih Zaki},
	month = sep,
	year = {2022},
	pmid = {35941929},
	pmcid = {PMC9348187},
	pages = {1433--1446},
}

@article{hamdi_attention-based_2022,
	title = {Attention-based hybrid {CNN}-{LSTM} and spectral data augmentation for {COVID}-19 diagnosis from cough sound},
	volume = {59},
	issn = {1573-7675},
	url = {https://doi.org/10.1007/s10844-022-00707-7},
	doi = {10.1007/s10844-022-00707-7},
	language = {en},
	number = {2},
	urldate = {2025-10-29},
	journal = {Journal of Intelligent Information Systems},
	author = {Hamdi, Skander and Oussalah, Mourad and Moussaoui, Abdelouahab and Saidi, Mohamed},
	month = oct,
	year = {2022},
	pages = {367--389},
}

@misc{hamdi_autoencoders_2022,
	type = {A4},
	title = {Autoencoders and {Ensemble}-{Based} {Solution} for {COVID}-19 {Diagnosis} from {Cough} {Sound}},
	copyright = {https://rightsstatements.org/vocab/InC/1.0/},
	url = {https://oulurepo.oulu.fi/handle/10024/53537},
	language = {eng},
	urldate = {2025-10-29},
	journal = {47f5ca1a-24da-457a-9afc-8438d6774f44},
	author = {Hamdi, Skander and Moussaoui, Abdelouahab and Oussalah, Mourad and Saidi, Mohamed},
	month = oct,
	year = {2022},
	note = {Accepted: 2025-01-03T06:56:00Z
Publisher: Springer},
}

@article{aytekin_covid-19_2024,
	title = {{COVID}-19 {Detection} {From} {Respiratory} {Sounds} {With} {Hierarchical} {Spectrogram} {Transformers}},
	volume = {28},
	issn = {2168-2208},
	doi = {10.1109/JBHI.2023.3339700},
	language = {eng},
	number = {3},
	journal = {IEEE journal of biomedical and health informatics},
	author = {Aytekin, Idil and Dalmaz, Onat and Gonc, Kaan and Ankishan, Haydar and Saritas, Emine Ulku and Bagci, Ulas and Celik, Haydar and Cukur, Tolga},
	month = mar,
	year = {2024},
	pmid = {38051612},
	pmcid = {PMC11658170},
	pages = {1273--1284},
}

@Article{s23114996,
AUTHOR = {Pavel, Irina and Ciocoiu, Iulian B.},
TITLE = {COVID-19 Detection from Cough Recordings Using Bag-of-Words Classifiers},
JOURNAL = {Sensors},
VOLUME = {23},
YEAR = {2023},
NUMBER = {11},
ARTICLE-NUMBER = {4996},
URL = {https://www.mdpi.com/1424-8220/23/11/4996},
PubMedID = {37299721},
ISSN = {1424-8220},
DOI = {10.3390/s23114996}
}

@article{Orlandic2021,
  author    = {Orlandic, Lara and Teijeiro, Tomas and Atienza, David},
  title     = {The COUGHVID crowdsourcing dataset, a corpus for the study of large-scale cough analysis algorithms},
  journal   = {Scientific Data},
  year      = {2021},
  volume    = {8},
  number    = {1},
  pages     = {156},
  doi       = {10.1038/s41597-021-00937-4},
  publisher = {Nature Publishing Group}
}

@misc{coughvid_zenodo_2020,
  author    = {Orlandic, Lara and Teijeiro, Tomas and Atienza, David},
  title     = {{The COUGHVID crowdsourcing dataset: A corpus for the study of large-scale cough analysis algorithms}},
  month     = sep,
  year      = 2020,
  publisher = {Zenodo},
  version   = {1.0},
  doi       = {10.5281/zenodo.4048312},
  url       = {https://doi.org/10.5281/zenodo.4048312}
}

@article{kong2020panns,
  title={PANNs: Large-scale pretrained audio neural networks for audio pattern recognition},
  author={Kong, Qiuqiang and Cao, Yin and Iqbal, Turab and Wang, Yuxuan and Wang, Wenwu and Plumbley, Mark D},
  journal={IEEE/ACM Transactions on Audio, Speech, and Language Processing},
  volume={28},
  pages={2880--2894},
  year={2020},
  publisher={IEEE}
}

@inproceedings{gemmeke2017audio,
  title={Audio set: An ontology and human-labeled dataset for audio events},
  author={Gemmeke, Jort F and Ellis, Daniel PW and Freedman, David and Jansen, Aren and Lawrence, Wade and Moore, R Chris and Plakal, Manoj and Ritter, Marvin},
  booktitle={2017 IEEE International Conference on Acoustics, Speech and Signal Processing (ICASSP)},
  pages={776--780},
  year={2017},
  organization={IEEE}
}

@article{pedregosa2011scikit,
  title={Scikit-learn: Machine learning in Python},
  author={Pedregosa, Fabian and Varoquaux, Ga{\"e}l and Gramfort, Alexandre and Michel, Vincent and Thirion, Bertrand and Grisel, Olivier and Blondel, Mathieu and Prettenhofer, Peter and Weiss, Ron and Dubourg, Vincent and others},
  journal={Journal of Machine Learning Research},
  volume={12},
  number={Oct},
  pages={2825--2830},
  year={2011}
}

@inproceedings{paszke2019pytorch,
  title={PyTorch: An imperative style, high-performance deep learning library},
  author={Paszke, Adam and Gross, Sam and Massa, Francisco and Lerer, Adam and Bradbury, James and Chanan, Gregory and Killeen, Trevor and Lin, Zeming and Gimelshein, Natalia and Antiga, Luca and others},
  booktitle={Advances in Neural Information Processing Systems 32},
  pages={8026--8037},
  year={2019}
}

@inproceedings{mcfee2015librosa,
  title={librosa: Audio and music signal analysis in python},
  author={McFee, Brian and Raffel, Colin and Liang, Dawen and Ellis, Daniel PW and McVicar, Matt and Battenberg, Eric and Nieto, Oriol},
  booktitle={Proceedings of the 14th Python in Science Conference},
  volume={8},
  year={2015}
}

@inproceedings{mckinney2010data,
  title={Data structures for statistical computing in python},
  author={McKinney, Wes},
  booktitle={Proceedings of the 9th Python in Science Conference},
  volume={445},
  pages={51--56},
  year={2010}
}

@article{sharma2020coswara,
  title={Coswara—A database of breathing, cough, and voice sounds for COVID-19 diagnosis},
  author={Sharma, Neeraj and Krishnan, P and Kumar, R and Ramoji, S and Chetupalli, S R and Nirmala, R and Ghosh, P K and Seshadri, Sriram},
  journal={arXiv preprint arXiv:2005.10548},
  year={2020}
}

@inproceedings{orlandic2021coughvid,
  title={The COUGHVID dataset: A large-scale pre-processed and augmented dataset of coughs},
  author={Orlandic, Lara and Teijeiro, Tomas and Atienza, David},
  booktitle={Proceedings of the 22nd Annual Conference of the International Speech Communication Association (INTERSPEECH)},
  pages={3361--3365},
  year={2021}
}

@article{kingma2014adam,
  title={Adam: A method for stochastic optimization},
  author={Kingma, Diederik P and Ba, Jimmy},
  journal={arXiv preprint arXiv:1412.6980},
  year={2014}
}

@misc{kong2019audioset_tagging_cnn,
  author = {Kong, Qiuqiang and others},
  title = {audioset\_tagging\_cnn},
  year = {2019},
  publisher = {GitHub},
  journal = {GitHub repository},
  howpublished = {\url{https://github.com/qiuqiangkong/audioset_tagging_cnn}}
}

@misc{hodges2020audio-pann-train,
  author = {Hodges, Martin},
  title = {audio-pann-train: Training and fine-tuning PANNs},
  year = {2020},
  publisher = {GitHub},
  journal = {GitHub repository},
  howpublished = {\url{https://github.com/MartinHodges/audio-pann-train}}
}
\nocite{*} 
\end{document}